\documentclass[apjl]{emulateapj}

\pdfoutput=1

\usepackage{comment}
\usepackage{ifthen}
\usepackage{amsmath}
\usepackage{amsfonts}
\usepackage{amssymb}
\usepackage{multirow}
\usepackage{wasysym}
\usepackage{subfigure}
\usepackage{epsfig}
\usepackage{color}
\definecolor{linkcolor}{rgb}{0,0,0.5}
\usepackage[%
	colorlinks=true,	
	pdfpagelabels=true,
	hypertexnames=true,
	plainpages=false,
	naturalnames=true,
	pdftitle={Priors for Orbital Eccentricity},    
	pdfauthor={David M. Kipping},     
	urlcolor=linkcolor,     
	linkcolor=linkcolor,    
	citecolor=linkcolor,    
	filecolor=linkcolor,    
        ]{hyperref}
\newcommand{\logg}{$\log g$}
\newcommand{\rhostar}{$\rho_{\star}$}
\newcommand{\Teff}{$T_{\mathrm{eff}}$}
\newcommand{\FeH}{[Fe/H]}
\newcommand{\Kepler}{\textit{Kepler}}

\newboolean{emulateapj}
\setboolean{emulateapj}{true}

\newboolean{astroph}
\setboolean{astroph}{true}

\newboolean{png}
\setboolean{png}{false}

\newboolean{color}
\setboolean{color}{true}

\shortauthors{Angus \& Kipping}
\shorttitle{Probabilistic Inference of Stellar Parameters}
\ifthenelse{\boolean{emulateapj}}{
    \newcommand{\titledag}{$\dagger$}
}{
    \newcommand{\titledag}{\dagger}
}

\begin{document}

\title {Probabilistic Inference of Basic Stellar Parameters:\\
Application to Flickering Stars 
\altaffilmark{\titledag}}

\author{
	{\bf	Ruth~Angus\altaffilmark{1},
		David.~M.~Kipping\altaffilmark{2},
	}
}

\altaffiltext{1}{Dept. of Physics, University of Oxford, UK; email:
		 ruth.angus@astro.ox.ac.uk}

\altaffiltext{2}{Department of Astronomy, Columbia University, 550 W 120th St.,
                 New York, NY 10027, US}

\altaffiltext{$\dagger$}{
Based on archival data of the \Kepler\ telescope.
}


\begin{abstract}

The relations between observable stellar parameters are usually assumed to be
deterministic.
That is, given an infinitely precise measurement of independent variable,
`$x$', and some model, the value of dependent variable, `$y$' can be known
exactly.
In practice this assumption is rarely valid and intrinsic stochasticity means
that two stars with exactly the same `$x$', will have slightly different
`$y$'s.
The relation between short-timescale brightness fluctuations (flicker) of
stars and both surface gravity \citep{bastien:2013} and stellar density
\citep{kipping:2014} are two such stochastic relations that have, until now,
been treated as deterministic ones.
We recalibrate these relations in a probabilistic framework, using
Hierarchical Bayesian Modelling (HBM) to constrain the intrinsic scatter in
the relations.
We find evidence for additional scatter in the relationships, that cannot be
accounted for by the observational uncertainties alone.
The scatter in surface gravity and stellar density does not depend on flicker,
suggesting that using flicker as a proxy for $\log g$ and $\rho_\star$ is
equally valid for dwarf and giant stars, despite the fact that the
observational uncertainties tend to be larger for dwarfs.

\end{abstract}



\section{INTRODUCTION}
\label{sec:intro}


Accurate stellar characterization plays a vital role for many active research
fields within astronomy. For example, stellar populations, galactic
archaeology, the study of binary stars, asteroseismology and exoplanet studies
all rely on inferences of basic stellar parameters to varying degrees.
Empirically-derived and reliable estimates are of particular value, increasing
our confidence in the end-product results built upon these inputs.

Basic stellar parameters, such as effective temperature and surface gravity,
can be inferred using one (or more) of several types of observations, such as
spectroscopy, photometry, interferometry, etc. This inference can be performed
by invoking theoretical models or by building an empirical calibration
library.
For example, an observed stellar spectrum could be matched against a library of
theoretical spectra generated using stellar atmosphere models, or, against a
library of observed spectra of ``standard stars'', serving as calibrators.
Regardless of the approach, be it theoretical or empirical, the methods used
for the inference of stellar parameters are traditionally ``deterministic''.
In this context, a deterministic model can be loosely described as one where
a particular observational input always returns a single-valued output for a
parameter of interest, i.e. nature itself has no variance and the underlying
model is considered to be a perfect description of reality.

An alternative approach for inferring model parameters is to allow
relationships between observables to be stochastic.
In recent years, there has been a shift towards such methods in several areas
of astronomy, particularly within the exoplanet community.
For example, \citet{wolfgang:2015} considered that the mass-radius
relationship of exoplanets is stochastic, since a particular sized planet
could be have a range of planet masses due to unmodeled variances in
compositions, environment and other complications.
These recent demonstrations in exoplanetary science have prompted us to
consider the need for treating the parent stars in the same probabilistic
framework, with potential applications spanning many fields of astronomy.

The demand for probabilistic stellar parameters is not only motivated by the
fact that probability distributions are far more representative of our
`beliefs' about astrophysical parameters, it also has a practical purpose.
When using data published in the astronomical literature to, for example,
infer relationships between parameters that are themselves the product of an
inference process (for example, exoplanet transit depth and period), inference
can be performed as the final stage in a hierarchical treatment \citep[see,
e.g.][]{foreman-mackey:2014}.
Studies such as these are benefited by posterior PDF samples, rather than
point estimates of inferred properties.



One of the more recent tools developed to characterize stars is known as
``flicker'' \citep{bastien:2013}.
Flicker is a proxy for the scatter on an 8-hour timescale (denoted as $F_8$)
in a broad visible bandpass time series photometric light curve, such as that
from \textit{Kepler} or the upcoming TESS mission. A more detailed account of
the proceedure to calculate flicker is described in \citet{bastien:2013}. As
shown in \citet{bastien:2013}, flicker displays a remarkable correlation to
the asteroseismically determined parent star surface gravities (\logg).
Turning this around, the observation implies that flicker can be used to
empirically infer surface gravities at the level of $\sim0.1$\,dex, an
attractive proposition given the wealth of photometric light curves available
through the array of exoplanet transit missions flying and scheduled to
launch.

\citet{cranmer:2014} demonstrated that models of stellar surface granulation
indeed reproduce a flicker effect in close agreement with that observed by
\citet{bastien:2013}, providing a physically-plausible explanation.
Since surface gravity is highly correlated with mean stellar density (\rhostar)
on evolutionary tracks, \citet{kipping:2014} showed that flicker can be also
be used to infer \rhostar, which is more useful for exoplanet transit analysis
\citep{seager:2003}.

Whether one calibrates flicker to \logg\ or \rhostar, there are several aspects
of the problem which are attractive for our purposes of a simple demonstration
of probabilistic inference of stellar parameters.
Firstly, in log-log space the relationship is very simple, appearing to be
linear \citep{kipping:2014}.
Secondly, there is a sufficiently large number of points in the sample (439
stars) to constrain a population-based model.
Thirdly, there is significant excess scatter around the best-fitting relation
implying that a deterministic model is inadequate.
This is not surprising given that granulation is a complex and messy process
for which one should not expect any parametric model to provide a perfect
description.
Finally, the physical processes that produce surface granulation, of which
flicker is an observational tracer, may be more or less noisy for different
types of stars.
We will test whether flicker has greater predictive power in certain regions
of parameter space; i.e. is flicker significantly more informative for
subgiants than for dwarfs?
For these reasons, we identify the calibration of
flicker to \logg\ and \rhostar\ as a well-posed problem to first demonstrate
probabilistic inference in the arena of stellar characterization.

\section{PROBABILISTIC CALIBRATION}
\label{sec:HBM}


\subsection{Calibration Data}

For our calibration data, we used a sample of \Kepler\ stars with
both asteroseismic and flicker measurements available. \citet{chaplin:2014}
report asteroseismic \rhostar\ estimates (and the associated uncertainties) for
518 \Kepler\ stars. The authors report three different sets of results,
depending on the choice of \Teff\ and \FeH, and in this work we elected to use
values reported in their Table 6 over Table 5, and Table 5 over Table 4. We
additionally used the 71 additional planet hosting stars with asteroseismology
reported in \citet{huber:2013} but not reported in \citet{chaplin:2014}. Values
for flicker and ``range'' were taken from \citet{kipping:2014}, based upon the
methods described in \citet{bastien:2013}.
In order to use the same data set as \citet{kipping:2014} and
for reasons described there-in, we only include targets in our calibration for
which:

\begin{itemize}
\item[{\tiny$\blacksquare$}] Range (defined in \citealt{bastien:2013})
$<1000$\,ppm
\item[{\tiny$\blacksquare$}] $4500<T_{\mathrm{eff}}<6500$\,K
\item[{\tiny$\blacksquare$}] $K_P<14$
\item[{\tiny$\blacksquare$}] $1.2 < \log_{10}$($F_8$\,[ppm])$< 2.2$
\end{itemize}

We use the same sample for our calibration of \logg, except that we exclude the
\citet{huber:2013} data, since these authors do not provide estimates of
\logg\footnote{Whilst we could compute \logg\ ourselves from the reported
masses and radii, this could only be done under the incorrect assumption of
zero covariance between $M_{\star}$ and $R_{\star}$.}.

\subsection{Hierarchical Bayesian Model}

We model the stochastic relationship between $F_8$, \logg\ and \rhostar,
accounting for the fact that there exists some intrinsic scatter in
the dependent variable.
There are two excellent reasons for modelling the relation stochastically;
firstly, if the intrinsic scatter is ignored and the relation between
variables is assumed to be deterministic, those data points with smaller
measurement uncertainties may have an unrepresentative greater weighting
during the fitting process \citep{hogg:2010b}.
Secondly, we are interested in producing probability distributions over
stellar densities and surface gravities, as opposed to point estimates, and
propagating these probability distributions through to subsequent analyses.
Several recent studies have required posterior Probability Density
Function (PDF) samples, in order to conduct their hierarchical analyses
\citep[e.g.][]{foreman-mackey:2014, rogers:2015, angus:2015}


The two models we use to describe the relationships between $F_8$, \logg\ and
\rhostar\ are
\begin{equation}
	\log_{10}(F_8) \sim \mathcal{N} \left(\alpha_\rho +
    \beta_\rho \log_{10}(\rho_\star), \sigma_\rho^2 \right),
\end{equation}
\label{eq:rho}
and
\begin{equation}
	\log_{10}(F_8) \sim \mathcal{N} \left(\alpha_g + \beta_g
    \log_{10}(g), \sigma_g^2 \right).
\end{equation}
\label{eq:logg}
The free parameters of the two models are $\alpha_\rho$, $\beta_\rho$,
$\sigma_\rho$, $\alpha_g$, $\beta_g$ and $\sigma_g$.
These relations are Gaussian distributions with means given by the equation of
a straight line and standard deviations which describe the intrinsic scatter
about the mean.
We used the MCMC package, {\tt emcee} \citep{foreman-mackey:2013} to explore the
posterior PDFs of our model parameters.

We also tested a model in which the additional scatter depends on flicker
itself, defined as
\begin{equation}
	\log_{10}(F_8) \sim \mathcal{N} \left(\mu = \alpha_\rho +
    \beta_\rho \log_{10}(\rho_\star), \sigma_{\rho}^2 + \gamma_\rho
    \log_{10}(F_8)
    \right),
\end{equation}
\label{eq:rho2}
for flicker vs $\rho_\star$ and similarly for \logg.
This model allowed us to determine whether there the magnitude of additional
scatter varied as a function of flicker.
In other words, whether flicker was a better proxy for \logg\ or \rhostar\ for
either dwarf or giant stars.
We found that the maximum {\it a-posteriori} values for the $\gamma$
parameters were consistent with zero: $\gamma_{rho} = 0.006 \pm 0.02, \gamma_g
= -0.01 \pm\ 0.01$, and interpret this as evidence for a constant intrinsic
scatter level across evolutionary stages.

We used a likelihood function which accounts for 2-D uncertainties but does
not allow the intrinsic scatter to be a function of the dependent or
independent variables.
For the relation between flicker and \rhostar, this likelihood function can be
written as
\begin{eqnarray}
	& \ln\left[p(F_8|\rho_\star, \alpha_\rho, \beta_\rho, \sigma_{\rho})\right]
        \propto  \\ \nonumber
    & -\frac{1}{2}\sum_{n=1}^N \left[\frac{[F_{8n}-(\alpha_\rho
    + \beta_\rho \rho_{*n})]^2}
	{\left[\beta_\rho \sigma_{F8, n}^2 + \sigma_{\rho *, n}^2 +
    \sigma_{\rho}^2\right]} + \ln(\sigma_{F8, n}^2) + \ln(\sigma_{\rho*, n}^2)
    + \ln(\sigma_\rho^2) \right]
	\\ \nonumber
\end{eqnarray}
\label{eq:likelihood}
and similarly for \logg.
We found that the posterior PDFs for the model parameters obtained using this
likelihood function were consistent with those obtained using a model that
only accounts for the uncertainties on the flicker measurements.
The median values of the model parameters differed by around $0.05
\sigma$ for the $\alpha$ and $\beta$ parameters, by $0.3 \sigma$ for
$\sigma_\rho$ and by $0.8 \sigma$ for $\sigma_g$.
Since they are so dependent on the observational uncertainties, the parameters
that describe the intrinsic scatter in the relations are more sensitive to
whether the uncertainties in the $x$-direction are included.
Accounting for uncertainties on $y$ {\it and} $x$ is not essential in this
case but is still good practice and will, in general, produce more accurate
model parameters and uncertainties.


We used the uninformative prior for the parameters of a straight line for
data with unknown uncertainties, outlined in \citet{vanderplas},
\begin{equation}
p(\alpha, \beta, \sigma) \propto \frac{1}{\sigma} \left( 1 + \beta^2
\right)^{-3/2}.
\end{equation}
\label{eq:priors}

\begin{figure}
\begin{center}
\includegraphics[width=8.4cm,angle=0,clip=true]{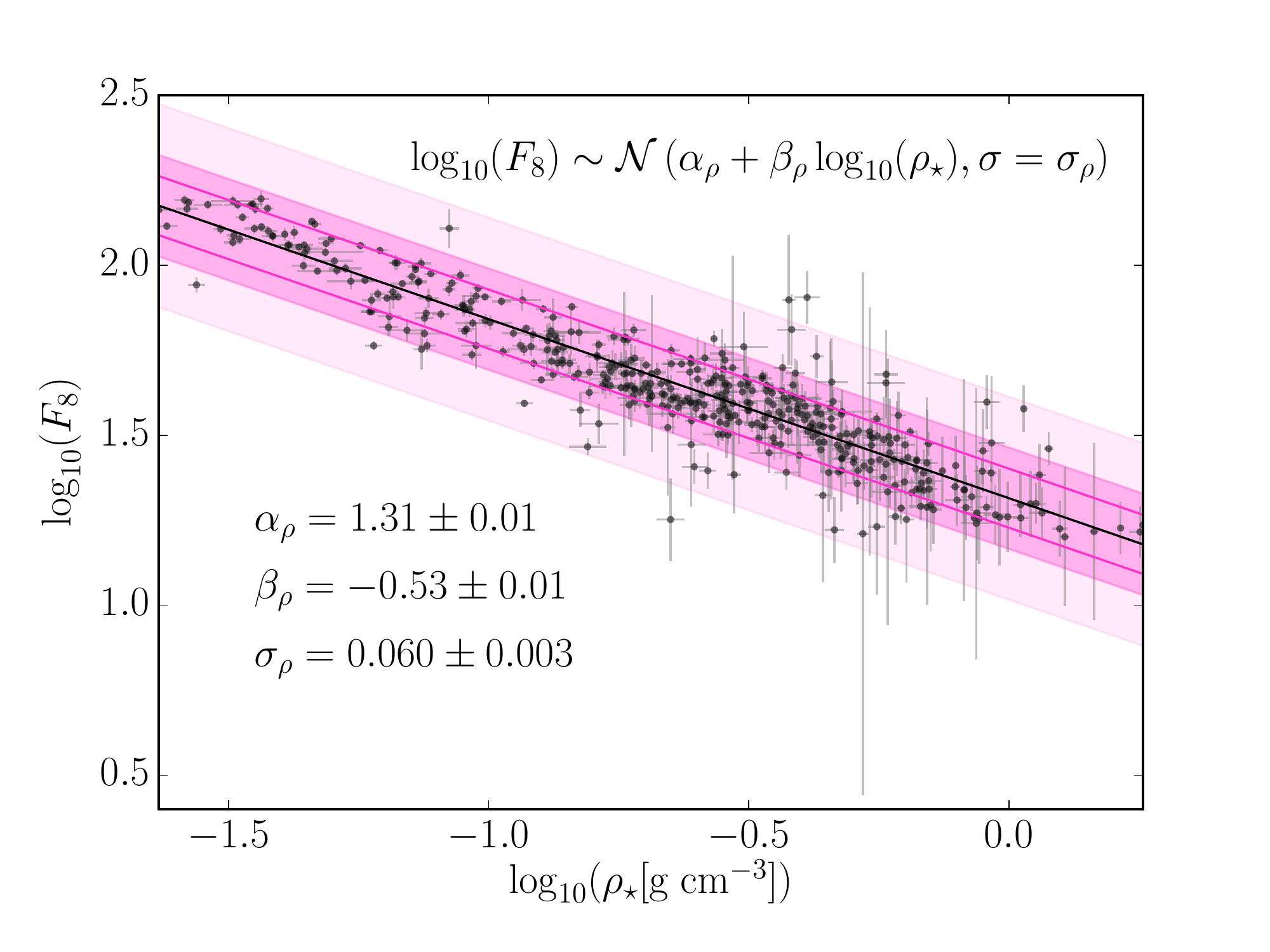}
\caption{
Stellar density vs. flicker.
This figure shows the model, conditioned on the data.
The solid black line shows the model with the best-fitting parameter values
quoted in the text.
The solid pink lines show the 1$\sigma$ region where the extra scatter is not
included and the pink shaded regions show the 1 and 2$\sigma$ regions with the
additional scatter.}
\label{fig:rhostar}
\end{center}
\end{figure}
\begin{figure}
\begin{center}
\includegraphics[width=8.4cm,angle=0,clip=true]{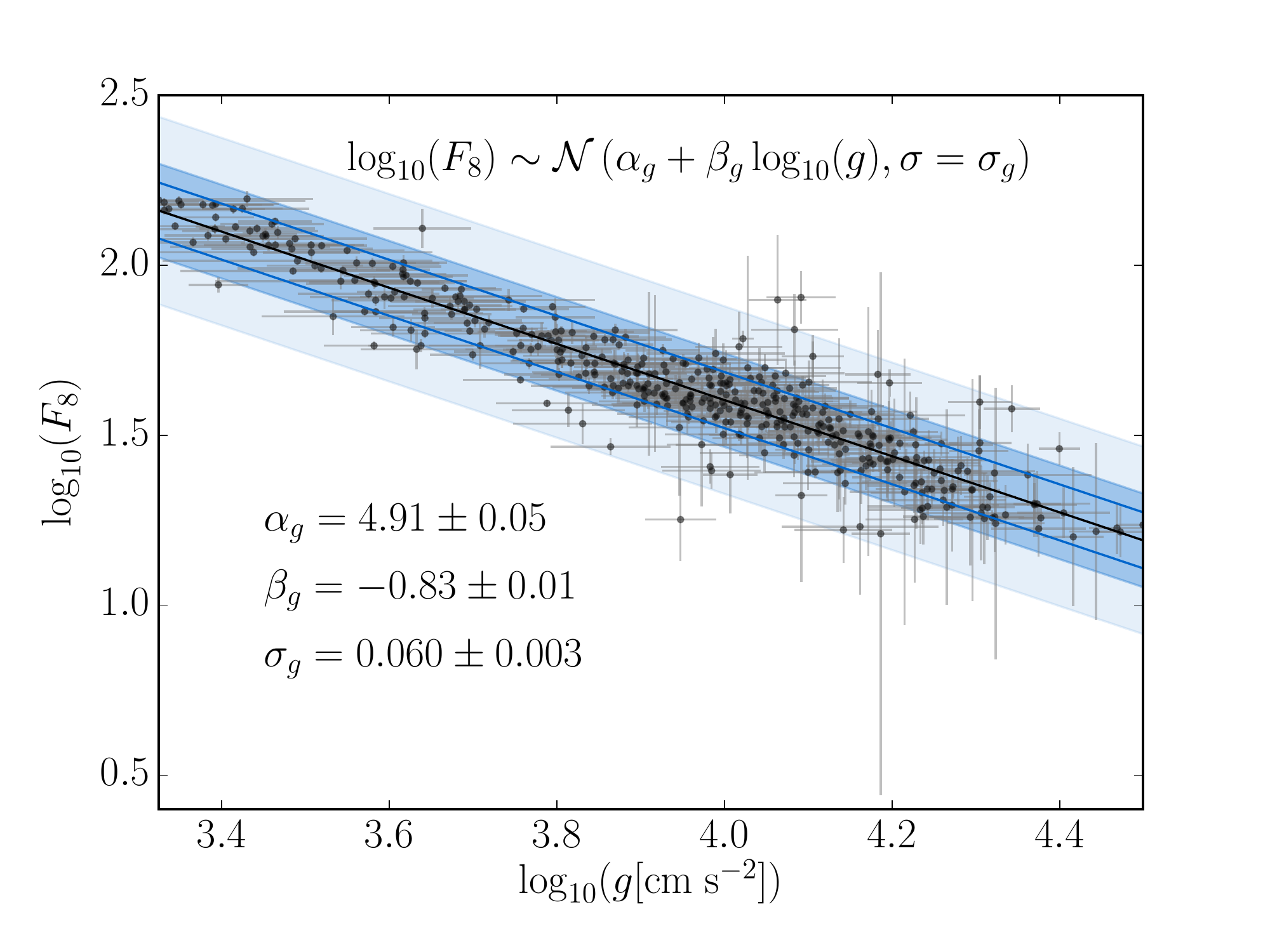}
\caption{
$\log(g)$ vs. flicker.
As in \ref{fig:rhostar} this figure shows the model, conditioned on the data.
The solid black line shows the model with the best-fitting parameter values
quoted in the text.
The solid blue lines show the 1$\sigma$ region where the extra scatter is not
included and the blue shaded regions show the 1 and 2$\sigma$ regions with the
additional scatter.}
\label{fig:logg}
\end{center}
\end{figure}
\begin{figure}
\begin{center}
\includegraphics[width=8.4cm,angle=0,clip=true]{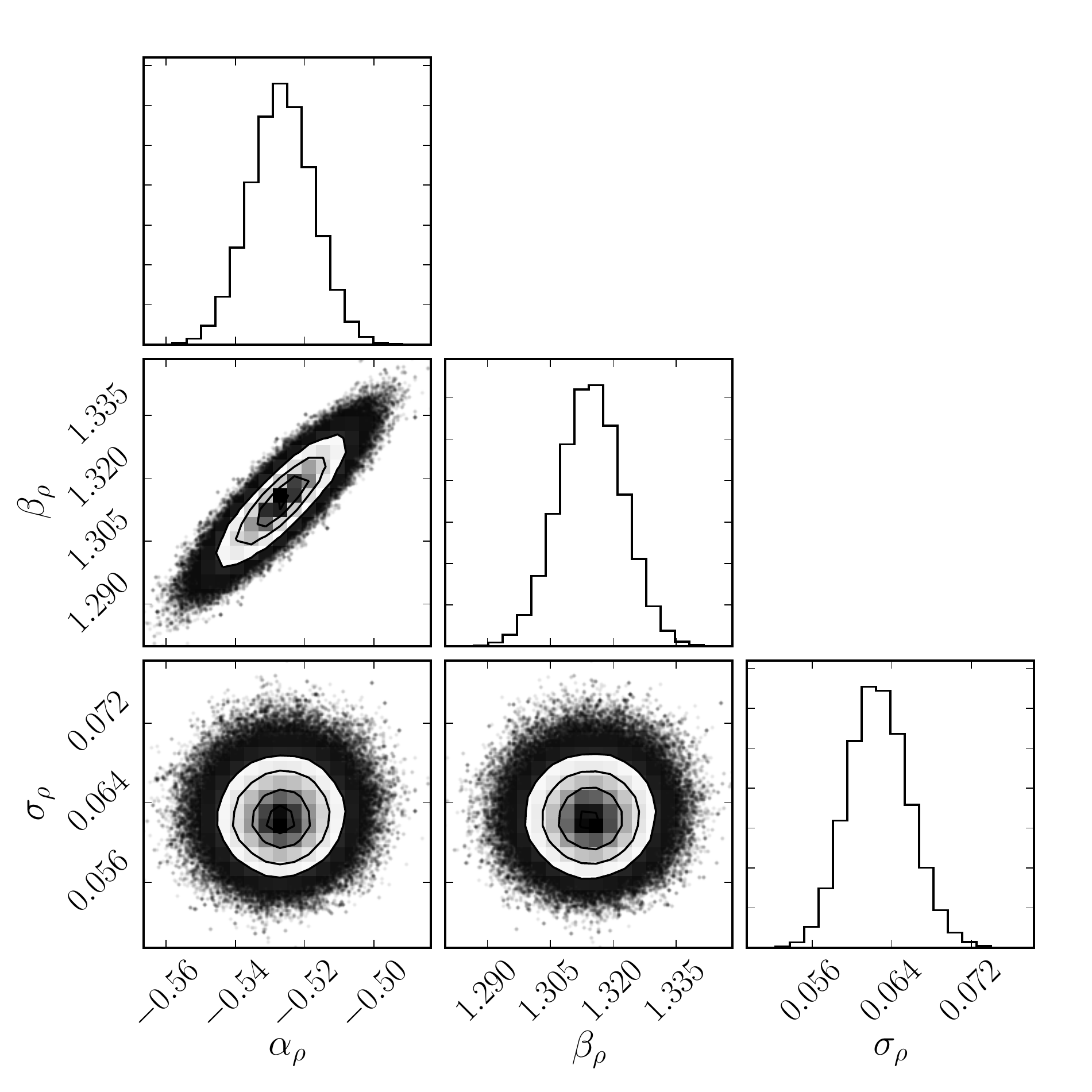}
\caption{Marginal posterior PDFs of the model parameters for \rhostar.
This figure was generated using the {\tt corner} python package
\citep{corner}.}
\label{fig:triangle}
\end{center}
\end{figure}

We also tested uniform, flat priors as defined below:
\begin{eqnarray}
 	&	\alpha, \beta \sim U(-10:10) \\ \nonumber
  	&	\log (\sigma_{rho}), \log(\sigma_g) \sim U(-10:10).
 \end{eqnarray}
 \label{eq:priors}
We found that the results were relatively insensitive to the choice of prior,
with median parameter values differing by only around $0.05\sigma$.
MCMC chains were run until the Gelman \& Rubin convergence criterion,
$\hat{R}$ reached a value of less than 1.002 and the number of autocorrelation
times was greater than 35.
Figures \ref{fig:rhostar} and \ref{fig:logg} show the data with the best-fit
models.
The shaded regions show the 1 and 2$\sigma$ confidence interval which are
representative of the intrinsic scatter in the relations.
The marginal posterior PDFs of the model parameters for \rhostar are
shown in figure \ref{fig:triangle}.
The marginal posterior PDFs for \logg\ are similarly Gaussian and, as with
$\sigma_\rho$, $\sigma_g$ is clearly greater than zero.
We checked the consistency between the two relations by calculating flicker
values for the Sun, finding $F_8 = 1.24 \pm 0.07$ and $F_8 = 1.21 \pm 0.1$
from Solar density and surface gravity measurements, respectively.
All the code used for this project and several ipython notebooks explaining
our analysis are available at {\url https://github.com/RuthAngus/flicker}.

\begin{table}
\begin{center}
\caption{Median parameter values with 1$\sigma$ uncertainties.}
\begin{tabular}{lcc}
\hline\hline
Parameter & Median value \\
    \hline
$\alpha_\rho$   &    1.31$\pm 0.01$   \\
$\beta_\rho$    &    -0.53$\pm 0.01$   \\
$\sigma_\rho$   &    0.060$\pm 0.003$   \\
\hline
$\alpha_g$      &    4.91$\pm 0.05$   \\
$\beta_g$       &   -0.83$\pm 0.01$   \\
$\sigma_g$      &    0.060$\pm 0.003$   \\
    \hline
\end{tabular}
\end{center}
\end{table}
\label{tab:results}

\section{DISCUSSION}
\label{sec:discussion}


We have recalibrated the relation between short timescale brightness
fluctuations in the {\it Kepler} light curves of stars (flicker) with both
stellar density and surface gravity, whilst including parameters to describe
the intrinsic scatter in these relationships, presented in table
\ref{tab:results}.
The terms $\sigma_\rho$ and $ \sigma_g$ are both non-zero, suggesting that
there {\it is} an additional source of scatter in the relations, not accounted
for by the observational uncertainties alone.
This is either caused by intrinsic scatter in the physical relationship
between flicker and density and \logg, produced by some physical process that
is not accounted for in the model, or by an underestimation of the
observational uncertainties.
We also tested a model with both an additional variance term {\it and} a term
that included flicker-dependent variance.
We found that the need for additional flicker-dependent variance was not
supported by the data, indicating that the intrinsic scatter in the relations
between flicker, \logg\ and \rhostar\ does not depend on evolutionary state.

This is a simple `fitting a line to data' exercise, however it continues the
discussion of probabilistic modelling that is an active topic within the
fields of exoplanet and stellar astronomy.
We used Hierarchical Bayesian Modelling (HBM) to constrain the intrinsic
scatter in the relationship between flicker, surface gravity and density and
included the effects of the non-negligible two-dimensional observational
uncertainties.
Relationships between astronomical parameters are almost always
non-deterministic; an element of stochasticity effects the physical parameters
of stars so one can never perfectly predict $y$ given an observation of $x$.
We advocate a probabilistic approach in both the `fitting the model to data'
step, {\it and} when using an empirically calibrated model to predict
parameter values.
The fitting stage benefits because if the relationships between parameters are
falsely assumed to be deterministic, they will be skewed by data points with
uncertainties that only represent measurement error and no additional scatter.
The prediction stage benefits from the stochastic treatment both because a
probability distribution is in many ways more representative of an observation
than a point estimate, and because posterior PDF samples can be used in
subsequent studies (provided the prior used during the fitting process is
described).

We provide posterior PDF samples at \url{https://zenodo.org/deposit/105051/}.
Whenever a prediction for the surface gravity or density of a star is required,
for a given estimate of flicker, we recommend using these posterior samples
within the calculation of \rhostar or \logg\ and its (Monte Carlo)
uncertainty.
These posterior samples will naturally fold in the covariances between
parameters.
Simple analytical uncertainty propagation is only valid when uncertainties are
Gaussian and uncorrelated which is rarely true and certainly not the case when
the model is a straight line (the slope and intercept are alway correlated).
A flicker value with uncertainties (or even better: posterior PDF
samples), input into our model will result in a probability distribution over
stellar densities or surface gravities which reflects both the uncertainties
on the flicker measurement, the uncertainties on the model parameters {\it and}
the intrinsic scatter in the flicker-\rhostar-\logg\ relations.

\acknowledgements
\section*{Acknowledgements}

RA would like to thank Dan Foreman-Mackey for his extremely helpful suggestions
and comments, Angie Wolfgang for her expert guidance and John Johnson for his
fantastic advice and support.
RA would also like to thank the anonymous referee for their patience and
excellent suggestions which dramatically improved this letter, the SAMSI
institute and the other members of the SAMSI {\it Kepler} exoplanet statistics
working groups.



\end{document}